\documentclass[prx,twocolumn,floatfix,superscriptaddress,amsmath,citeautoscript,aps,longbibliography]{revtex4-1}
\usepackage{blindtext}
\usepackage{graphicx}
\usepackage{tabularx}
\usepackage{array}
\usepackage{booktabs}
\usepackage{dcolumn}
\usepackage{bm}
\usepackage{color}
\usepackage[normalem]{ulem}
\usepackage{natbib}
\newcommand{\stkout}[1]{\ifmmode\text{\sout{\ensuremath{#1}}}\else\sout{#1}\fi}
\definecolor{magenta}{cmyk}{ 0, 1, 0,0}
\usepackage{amsmath}
\usepackage{float}
\usepackage{bm}
\usepackage{wasysym}
\usepackage{empheq}

\usepackage{hyperref}  
\hypersetup{hypertex=true,
            colorlinks=true,
            linkcolor=blue,
            anchorcolor=blue,
            citecolor=blue}

\begin{document}
\title{Magnetic structure and Ising-like antiferromagnetism in the \\bilayer triangular lattice compound NdZnPO}

\author{Han Ge}
\affiliation{Department of Physics, Southern University of Science and Technology, Shenzhen 518055, China}

\author{Tiantian Li}
\affiliation{Department of Physics, Southern University of Science and Technology, Shenzhen 518055, China}

\author{S. E. Nikitin}
\affiliation{Laboratory for Neutron Scattering and Imaging, Paul Scherrer Institut, CH-5232 Villigen-PSI, Switzerland}

\author{Nan Zhao}
\affiliation{Department of Physics, Southern University of Science and Technology, Shenzhen 518055, China}

\author{Fangli Li}
\affiliation{Department of Physics, Southern University of Science and Technology, Shenzhen 518055, China}

\author{Huanpeng Bu}
\affiliation{Department of Physics, Southern University of Science and Technology, Shenzhen 518055, China}

\author{Jiayue Yuan}
\affiliation{Department of Physics, Southern University of Science and Technology, Shenzhen 518055, China}

\author{Jian Chen}
\affiliation{Department of Physics, Southern University of Science and Technology, Shenzhen 518055, China}

\author{Ying Fu}
\affiliation{Department of Physics, Southern University of Science and Technology, Shenzhen 518055, China}
\affiliation{Quantum Science Center of Guangdong-Hong Kong-Macao Greater Bay Area (Guangdong), Shenzhen 518045, China}

\author{Jiong Yang}
\affiliation{Department of Chemistry, Southern University of Science and Technology, Shenzhen 518055, China}

\author{Le Wang}
\affiliation{Shenzhen Institute for Quantum Science and Engineering, Southern University of Science and Technology, Shenzhen 518055, China}
\affiliation{International Quantum Academy, Shenzhen 518048, China}

\author{Ping Miao}
\affiliation{Institute of High Energy Physics, Chinese Academy of Sciences, Beijing 100049, China}
\affiliation{Spallation Neutron Source Science Center, Dongguan 523803, China}

\author{Qiang Zhang}
\affiliation{Neutron Scattering Division, Oak Ridge National Laboratory, Oak Ridge, Tennessee 37831, USA}


\author{In$\acute{\rm e}$s Puente-Orench}
\affiliation{Instituto de Nanociencia y Materiales de Arag$\acute{o}$n (INMA), CSIC-Universidad de Zaragoza, Zaragoza 50009, Spain}
\affiliation{Institut Laue-Langevin, 71 Avenue des Martyrs, CS 20156, CEDEX 9, 38042 Grenoble, France}

\author{Andrey Podlesnyak}
\affiliation{Neutron Scattering Division, Oak Ridge National Laboratory, Oak Ridge, Tennessee 37831, USA}

\author{Jieming Sheng}
\thanks{Corresponding author: shengjm@gbu.edu.cn}
\affiliation{School of Physical Sciences, Great Bay University, Dongguang 523000, China}
\affiliation{Great Bay Institute for Advanced Study, Dongguang 523000, China}

\author{Liusuo Wu}
\thanks{Corresponding author: wuls@sustech.edu.cn}
\affiliation{Department of Physics, Southern University of Science and Technology, Shenzhen 518055, China}
\affiliation{Quantum Science Center of Guangdong-Hong Kong-Macao Greater Bay Area (Guangdong), Shenzhen 518045, China}
\affiliation{Shenzhen Key Laboratory of Advanced Quantum Functional Materials
   and Devices, Southern University of Science and Technology, Shenzhen 518055, China}

\date{\today}
\begin{abstract}

The complex interplay of spin frustration and quantum fluctuations in low-dimensional quantum materials leads to a variety of intriguing phenomena. This research focuses on a detailed analysis of the magnetic behavior exhibited by NdZnPO, a bilayer spin-1/2 triangular lattice antiferromagnet. The investigation employs magnetization, specific heat, and powder neutron scattering measurements.
At zero field, a long-range magnetic order is observed at $T_{\rm N}=1.64~\rm K$. 
Powder neutron diffraction experiments show the Ising-like magnetic moments along the $c$-axis, revealing a stripe-like magnetic structure with three equivalent magnetic propagation vectors.
Application of a magnetic field along the $c$-axis suppresses the antiferromagnetic order, leading to a fully polarized ferromagnetic state above $B_{\rm c}=4.5~\rm T$. This transition is accompanied by notable enhancements in the nuclear Schottky contribution. Moreover, the absence of spin frustration and expected field-induced plateau-like phases are remarkable observations. Detailed calculations of magnetic dipolar interactions revealed complex couplings reminiscent of a honeycomb lattice, suggesting the potential emergence of Kitaev-like physics within this system. This comprehensive study of the magnetic properties of NdZnPO highlights unresolved intricacies, underscoring the imperative for further exploration to unveil the underlying governing mechanisms.

\end{abstract}
\maketitle
\section{Introduction}

Geometrically frustrated quantum magnets have drawn significant interest over the past few decades~\cite{QSL1,QSL2,QSL3,QSL4,QSL5}. The two-dimensional (2D) triangular antiferromagnetic (AFM) spin lattice serves as a prototype example among different models of frustrated spins. Subsequent theoretical and numerical studies have been conducted to reveal the ground state of the two-dimensional triangular antiferromagnet~\cite{TAFmodel1,TAFmodel2,TAFmodel3,CsCeSe2}.  Nevertheless, there are still many problems to be solved, and further efforts are necessary to address the challenges.

Previous studies have been focused on the $d$-electron based systems, such as the organic Mott insulator $\kappa$-(BEDT-TTF)$_2$Cu$_2$(CN)$_3$ with a nearly isotropic triangular lattice~\cite{k-BTCuCN1}, the orthorhombic perovskite compounds Cs$_2$CuBr$_4$ with an anisotropic triangular lattice~\cite{CsCuBr1}, and the Co-based compound Ba$_3$CoSb$_2$O$_9$ with an ideal equilateral triangular lattice~\cite{BaCoSbO1}. Many of these compounds have very high critical fields, making it challenging to carefully investigate the complete phase diagram through thermodynamic techniques and neutron scattering experiments~\cite{CsCuBr2,BaCoSbO2,BaCoSbO3}.
On the other hand, rare-earth based compounds typically exhibit relatively smaller critical fields due to the large effective $g$-factor arising from strong spin-orbit coupling~\cite{RE1,RE2}. In addition, the crystalline electrical field (CEF) usually results in a degenerate doublet ground state, which allows us to describe it by a Hamiltonian with an effective spin S$_{\mathrm{eff}}=1/2$~\cite{Pseudospin1,Pseudospin2}.
In recent years, significant focus has turned to rare-earth-based quantum magnets like RMgGaO$_4$ (R = Yb, Tm)~\cite{YMGO,TMGO} and AYbCh$_2$ (A=Na, K, Cs; Ch = O, S, Se)~\cite{NYO,NYS,NYSe,KYSe,CYSe}.
These materials have garnered attention due to the observation of various novel phenomena, such as Kosterlitz-Thouless phase transitions, broad continuum-like excitations featuring a spinon Fermi surface, quantum spin liquid ground state, field-tunable quantum disordered ground state, etc.

In addition to this, another example of achieving robust spin frustration within a two-dimensional lattice is the Kitaev model. This model involves interacting spins positioned on a honeycomb lattice, characterized by the bond-directional exchange interaction~\cite{Kitaev}. 
What makes this model particularly intriguing is its exact solution, offering the initial concrete demonstration of how a quantum spin liquid with fractionalized excitations can arise in a two-dimensional spin model. 
In recent years, many 4$d$ and 5$d$ materials with 2D honeycomb lattices have been explored, including the iridates, $\alpha$-Na$_2$IrO$_3$~\cite{Na2IrO3}, $\alpha$-Li$_2$IrO$_3$~\cite{Li2IrO3}, and H$_3$LiIr$_2$O$_6$~\cite{H3LiIr2O6}, and the halide $\alpha$-RuCl$_3$~\cite{RuCl3}. These compounds exhibit stronger spin-orbital couplings than 3$d$ systems, and they were believed to have Kitaev-like anisotropic exchange interactions~\cite{Na2IrO3,Li2IrO3,H3LiIr2O6,RuCl3}. Particularly, recent studies on the honeycomb halide compound $\alpha$-RuCl$_3$ reveal that it can approximately realize the Kitaev spin liquid phase in an intermediate field region between the low-field zigzag ordered state and the high-field polarized phase~\cite{RuCl3-2, RuCl3-3}. Additionally, the thermal Hall conductivity measurements indicate a half-integer quantized state in these intermediate fields. Although the precise nature of this state is still under debate, the underlying physics has become increasingly important.
Furthermore, the bilayer structure of the rare-earth isotropic triangular lattice compound YbOCl, where Yb$^{3+}$ ions can be considered as a honeycomb lattice, positions it as a promising candidate for probing the Kitaev quantum spin liquid~\cite{YOC1,YOC2}.

Motivated by these studies, we explored a spin lattice consisting of two magnetic triangular layers arranged in an A-B stacking configuration.
In this bilayer structure, the lattice sites of one triangular layer project to the centers of the other triangular plaquettes next to it. This is equivalent to a honeycomb lattice with nearest and next-nearest neighbor interaction, and each triangular layer can be treated as one of the two sublattices of the honeycomb lattice~\cite{Triangular_honeycomb}. The ground state of this bilayer triangular lattice is highly degenerated, and strong spin frustration and related phenomena are then expected. It was reported that the rare-earth based family compound RZnPO (R=rare earth) adopted a trigonal crystal structure, where the magnetic rare-earth ions form two-dimensional bilayer triangular lattices~\cite{RZPO1,RZPO2,RZPO3}. 
In this manuscript, our attention is directed towards the Neodymium (Nd) based compound NdZnPO. We observed a long-range stripe-like order below $T_{\rm N}=1.64$ K, and the magnetic structure of NdZnPO was thoroughly examined through powder neutron scattering experiments. 
We constructed a field-temperature phase diagram utilizing magnetization and specific heat measurements, extending our analysis to temperatures as low as 60~mK. Simultaneously, CEF analysis was conducted, uncovering an Ising-like magnetic ground state.


\begin{figure*}[hbt!]
  \includegraphics[width=1\textwidth]{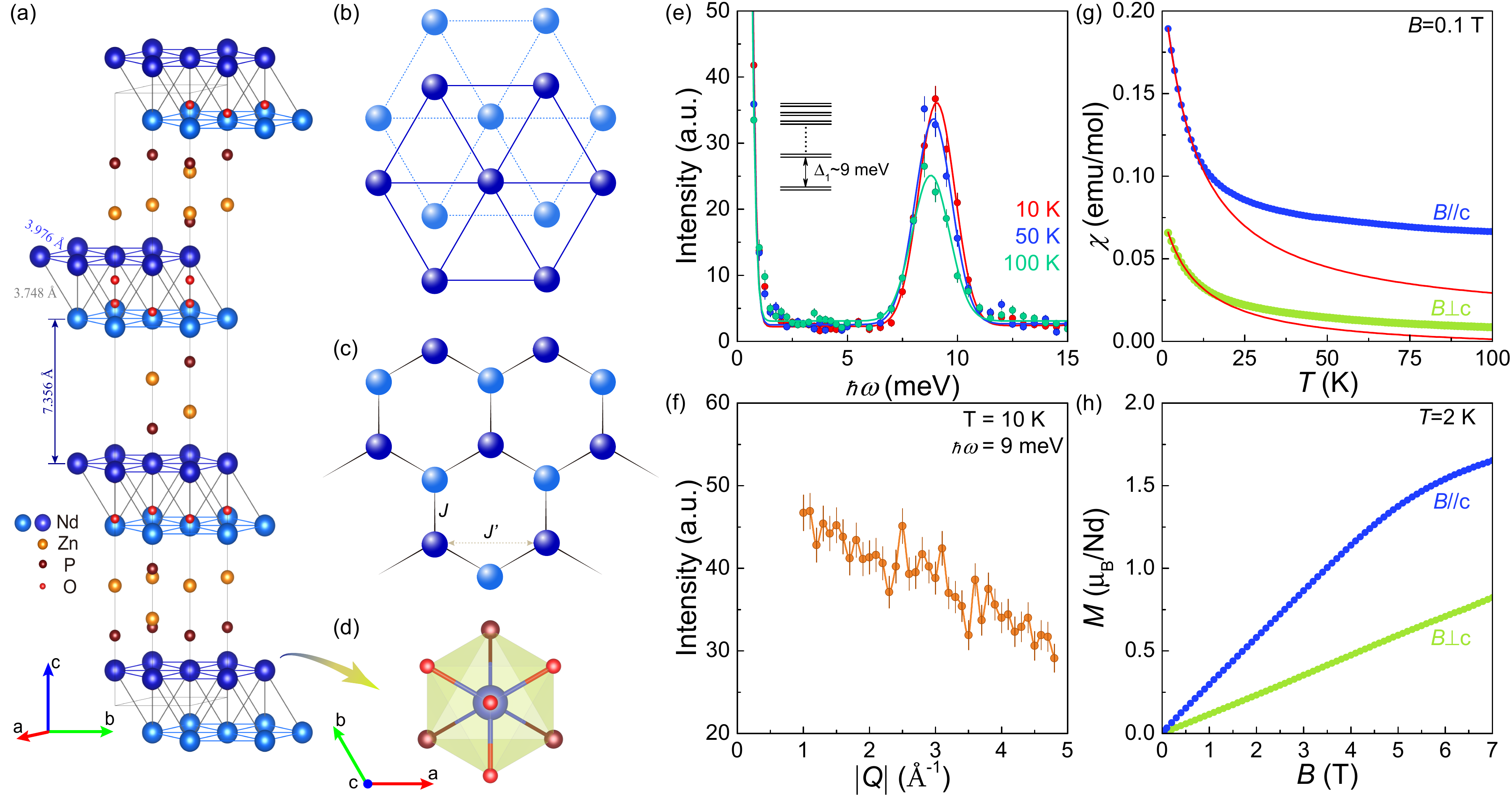}
  \caption{(a) Crystal structure of NdZnPO.
  (b) The bilayer triangular lattice of magnetic Nd$^{3+}$ ions in the $ab$ plane.
  (c) Representation of an equivalent honeycomb lattice structure illustrating the nearest-neighbor ($J$) and next-nearest neighbor ($J^{\prime}$) interactions. Each triangular layer serves as a sublattice within the honeycomb lattice.
  (d) Nd-P-O polyhedron with local point symmetry $C_{\rm 3v}(3m)$ at the Nd site.
  (e) Inelastic scattering over the wave vector $Q=3~\rm \AA^{-1}$ at different temperatures, demonstrating the first CEF level around 9.06 meV. The elastic peak and the first excited CEF peak have been fitted with a Gaussian function, as indicated by the solid lines. Inset: A scheme of five isolated CEF doublets.
  (f) Inelastic scattering intensity at 10 K as a function of the wave vector with a constant energy transfer of $\hbar \omega=9$ meV.
  (g) Temperature dependence of the magnetic susceptibility $\chi(T)$ measured in a field of 0.1 T along the $c$ direction and in the ab-plane. The solid red lines are the fits using the Curie-Weiss law.
  (h) Isothermal magnetization $M(B)$ measured at 2 K with the field applied along and perpendicular to the $c$ axis.
  }
\label{Structure_anisotropy}
\end{figure*}

\begin{figure}[hbt!]
 \includegraphics[width=0.48\textwidth]{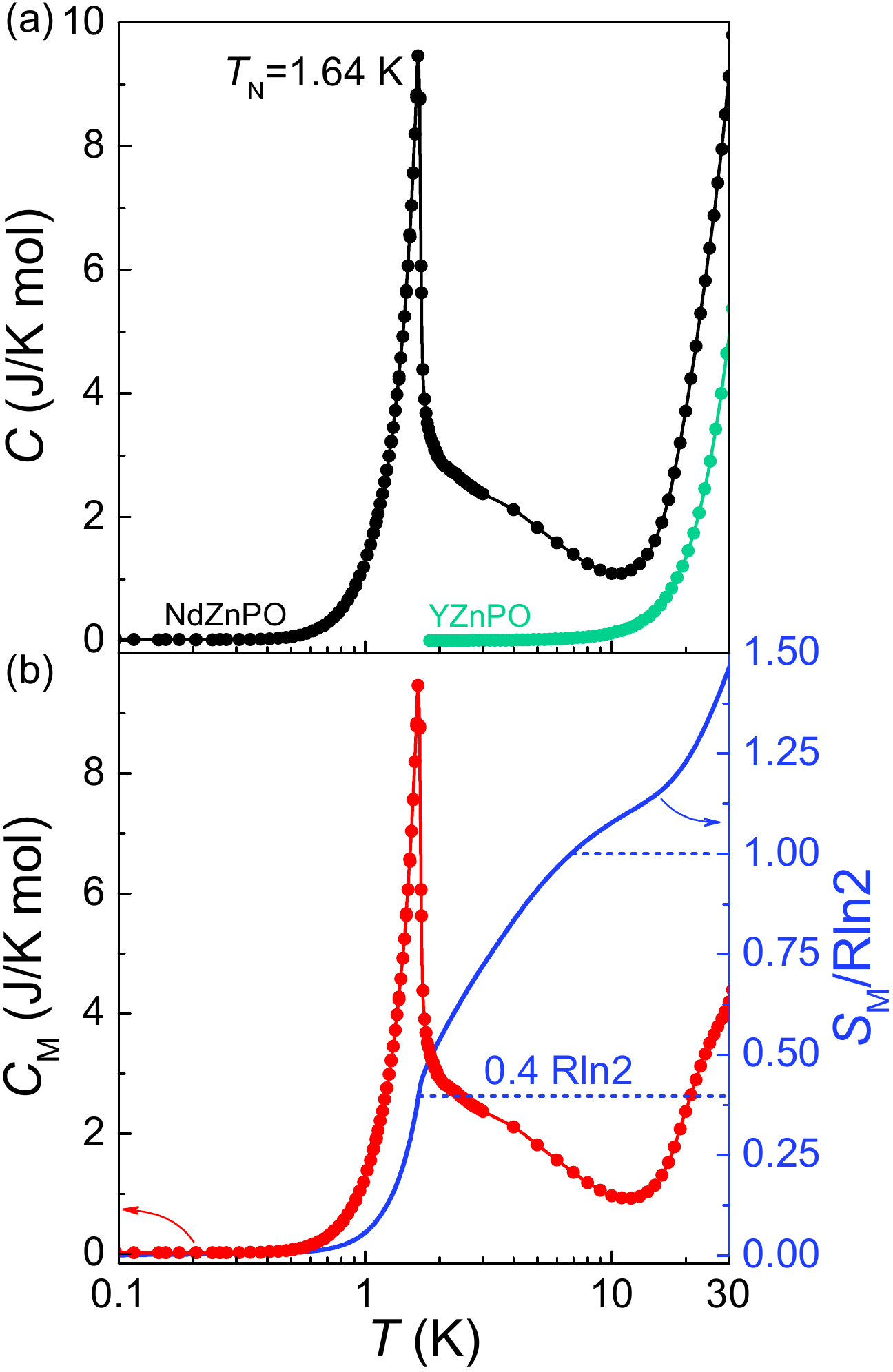}
    \caption{(a) Temperature-dependent specific heat $C$ of NdZnPO~(black circles, $C_{\rm Nd}$) and YZnPO~(green circles, $C_{\rm Y}$) in zero field. (b) Temperature dependence of the magnetic specific heat (red circles, $C_{\rm M}=C_{\rm Nd}-C_{\rm Y}$) with nonmagnetic contributions removed. The temperature-dependent normalized integrated magnetic entropy $S_{\rm M}$/Rln2 is shown on the right axis (blue line). A total entropy of $R$ln2 is released around 7 K.}
    \label{CT}
\end{figure}

\begin{figure*}[hbt!]
 \includegraphics[width=1\textwidth]{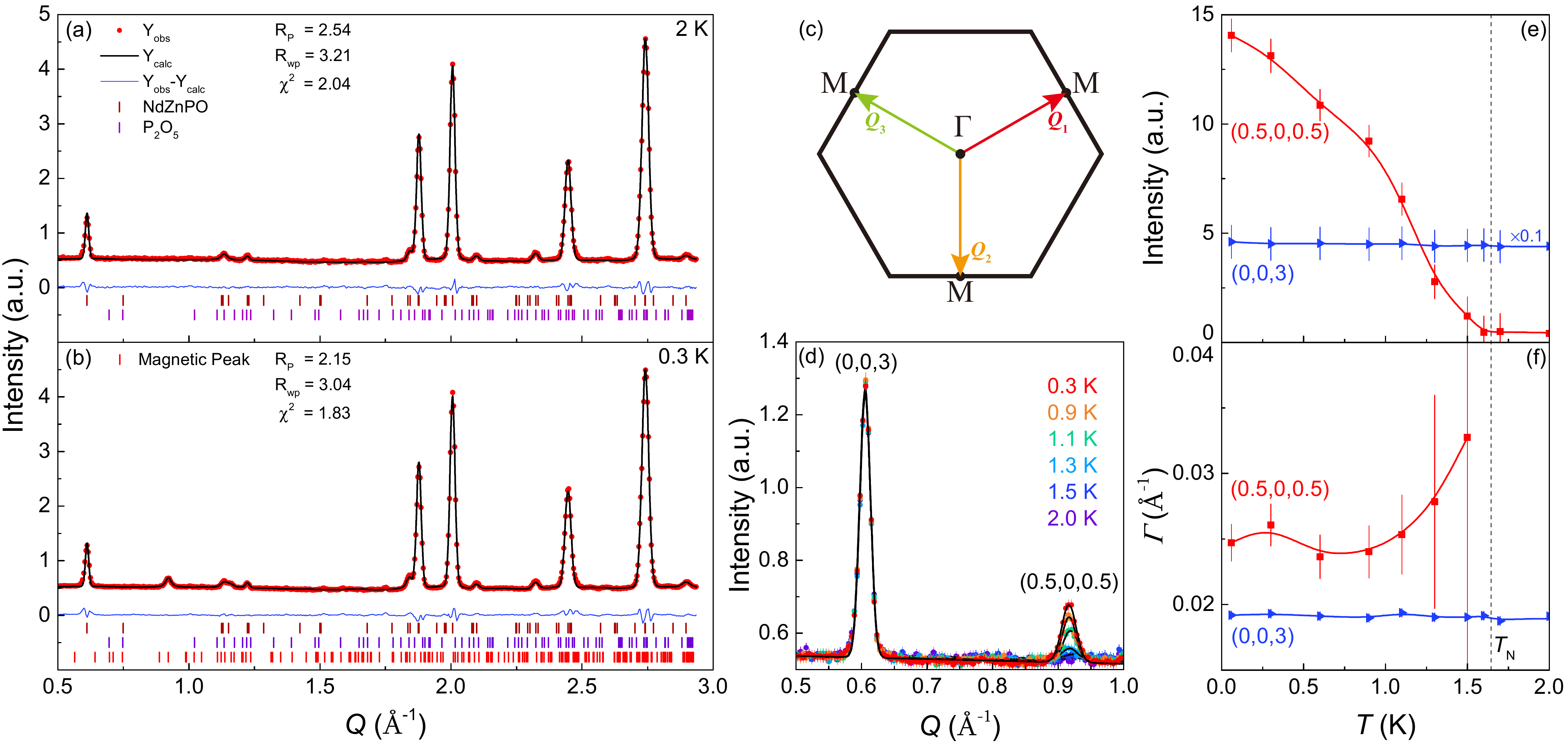}
    \caption{Rietveld refinement of neutron powder diffraction patterns for NdZnPO at (a) 2 K and (b) 0.3 K.
    The red points represent actual data, and the black lines represent the Rietveld fitting to the data. The difference curve is shown in the blue line. 
    (c) Schematic of three equivalent magnetic propagation vectors $\mathbf{Q_1}=(0.5,0,0.5)$, $\mathbf{Q_2}=(0,-0.5,0.5)$, and $\mathbf{Q_3}=(-0.5,0.5,0.5)$, from $\Gamma$ point to $\rm M$ point. The black hexagon represents the Brillouin zone in reciprocal space.
    (d) Neutron powder diffraction pattern of NdZnPO at different temperatures with the first two peaks (0,0,3) and (0.5,0,0.5). The circles represent actual data, and the black lines depict Gaussian fitting to the data. Temperature evolution of (e) the integrated intensity and (f) the fitted FWHM ($\Gamma$) of the two peaks.}
    \label{mag_peak}
\end{figure*}

\begin{figure*}[hbt!]
 \includegraphics[width=1\textwidth]{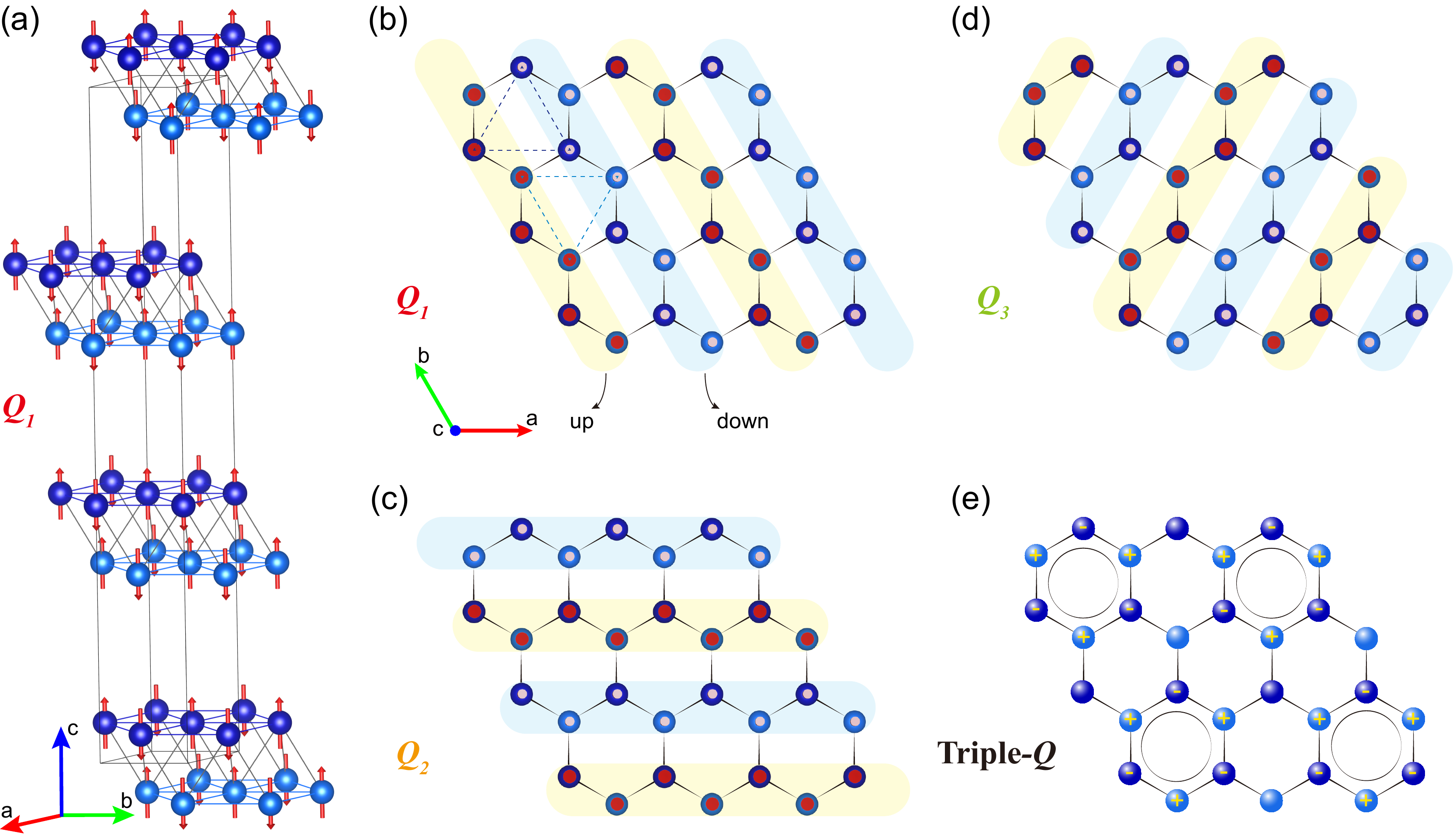}
    \caption{(a) The magnetic structure of NdZnPO below the ordering temperature refined from the neutron data. The red arrows indicate the ordered Nd spins. (b-d) The spin configuration of the bilayer triangular lattice of Nd$^{3+}$ ions projected onto the ab-plane with the magnetic propagation vectors~$\mathbf{Q_1}$, $\mathbf{Q_2}$, and $\mathbf{Q_3}$, respectively. The regular triangle in solid lines and the inverted triangle in dotted lines represent the upper and lower triangular lattice layers respectively. The yellow/blue bar represents the spin-up and spin-down respectively.
    (e) Possible triple-$Q$ spin configuration in the honeycomb lattice. The yellow symbol +/- represents spin-up and spin-down respectively.}    
    \label{mag_structure}
\end{figure*}

\section{Experimental method}

NdZnPO single crystals were grown using the flux method~\cite{RZPO1}. High-purity starting materials, including zinc oxide [ZnO, Aladdin, 99.999\%], neodymium [Nd, Alfa Aesar, 99.9\%], and phosphorus [P, Aladdin, 99.999\%], were mixed in an alumina crucible with NaCl/KCl as the flux. The mixture was heated up to 900$^{\circ}\rm C$ and then slowly cooled down to 300$^{\circ}\rm C$ at a rate of 5$^{\circ}\rm C$/hour. The excess flux was removed with distilled water. Hexagonal-shaped NdZnPO single crystals with sizes around $0.2\times0.2\times0.05$ mm$^3$ were obtained. The crystal structure of NdZnPO was verified using the Bruker D8 VENTURE diffractometer. Measurements of magnetization, magnetocaloric effect (MCE), and specific heat were carried out using the commercial Quantum Design Magnetic Property Measurement System and the Physical Property Measurement System with the dilution refrigerator insert.

Inelastic neutron scattering experiment was performed at the Swiss spallation neutron source (SINQ), hosted at the Paul Scherrer Institute (PSI). The experiment utilized a thermal neutron triple-axis spectrometer EIGER with fixed $E_{\rm f}~=~14.68$~meV. A pyrolotic graphite filter was installed between the sample and the analyzer to suppress higher-order scattering. The powder sample with $m~\approx$~2~g, was enclosed in a 10~mm aluminum can and cooled down using an orange cryostat. 
Neutron powder diffraction experiment was carried out using the high-intensity two-axis powder diffractometer, D1B, at the Institut Laue-Langevin (ILL). The neutron wavelength was fixed at $\lambda=2.52$ {\AA}, with a full width at half maximum (FWHM) of $\sim$0.3$^{\circ}$~\cite{D1B}. The powder sample, $\sim$2~g in mass, was cooled down to 60~mK using a dilution refrigerator insert. The dataset is available from the ILL database~\cite{dataILL}.

\section{Results and discussions}
\subsection{Crystal structure, CEF and single-ion anisotropy}
\label{structure}
NdZnPO exhibits a trigonal crystal structure characterized by the space group $R\bar{3}\rm m$ (No.166), featuring refined lattice parameters of $a=b=3.976$~{\AA} and $c=30.955$~\AA~\cite{RZPO1}. As shown in  Fig.~\ref{Structure_anisotropy}(a), the magnetic Nd$^{3+}$ ions are located at the center of the Nd-P-O polyhedron, forming a bilayer triangular network in the crystal $ab$ plane. Within each triangular layer, the nearest neighbor distance is 3.976~{\AA}, while the nearest neighbor distance between adjacent layers is 3.748~{\AA}. Within this bilayer structure, the upper triangular layer is center-stacked on top of the lower triangular layer (Fig.~\ref{Structure_anisotropy}(b)), which is equivalent with a 2D honeycomb-like lattice with nearest-neighbor interactions ($J$) and next-nearest-neighbor interactions ($J^\prime$) from the topside view (Fig.~\ref{Structure_anisotropy}(c)). These bilayer triangular planes are then alternatively stacked with a `CA'`AB'`BC'`CA' type sequence along the c-axis. The non-magnetic Zn and P ions are stuffed in between these bilayer planes, resulting in well-separated double magnetic triangular layers with an inter-double layer distance of about 7.356~\AA.

Quantum fluctuations are usually more pronounced for magnetic ions with small spin values, such as $S = 1/2$. The angular momentum $J$ in rare-earth ions is often regarded as a `classical' moment due to its large angular moment. However, as these rare-earth ions form crystals, the situation can be quite different. In the local crystalline environment of NdZnPO, the magnetic Nd$^{3+}$ ions are surrounded by four nearby O$^{2-}$ ions and three P$^{4+}$ ions, resulting in a local point symmetry $C_{\rm 3v}(3m)$ (Fig.~\ref{Structure_anisotropy}(d)). In this local point symmetry, the degenerate $2J+1=10$ ($L=6, S=3/2$, and $J=9/2$) multiplet states of Nd$^{3+}$ split into five doublet CEF states. To describe these CEF states, calculations based on the point-charge model have been performed using the software package MCPHASE~\cite{MCPHASE1,MCPHASE2}. By choosing the local $z$-axis along the crystal $c$ axis, the calculated ground doublet states are
\begin{equation}\label{CEF}
|E_{0\pm}\rangle =0.955 |\pm 9/2\rangle \mp 0.268 |\pm 3/2\rangle + 0.128 |\mp 3/2\rangle.
\end{equation}
These ground doublet states are dominated by wave functions $|\pm 9/2\rangle$, suggesting an Ising-like character with moments mainly along the easy $c$-axis. The other four excited doublet states are located at 13.7, 19.6, 26.5, and 42.7 meV, respectively, which are all well separated from the ground doublet states. Thus, the magnetic properties are mainly governed by the ground doublet states at sufficiently low temperatures, and the low-temperature properties can be described using an effective model of spin $S_{\rm eff}=1/2$. However, we have to notice that, in the case of a nearly pure Ising-like ground state with wave functions (\ref{CEF}), the term that flips the upside and downside spins is negligibly small,
\begin{equation}\label{spin_flip}
\langle E_{0\mp}|S^{\pm}|E_{0\pm}\rangle\simeq0.
\end{equation}
This indicates that neutrons can neither flip these Ising-like Nd$^{3+}$ moments nor make them propagate, resulting in an effective Ising-like spin Hamiltonian. Similar situations have been found in the rare-earth perovskites, where spin waves or any excitations within the ground doublet states are missing~\cite{DSO,TSO}. However, the CEF excitation from the ground doublet to the first excited states is still allowed.

Figure~\ref{Structure_anisotropy}(e) displays the inelastic neutron spectra taken at $T = 10$, 50 and 100 K.
A single excitation centered around $\Delta_1 = 9.06$ meV was observed for all temperatures (Fig.~\ref{Structure_anisotropy}(e)), which is close to the location of the first excited states calculated from the point charge model. Additionally, the scattering intensity at this center position decreases with increasing the wave vector $|\mathbf{Q}|$, confirming the magnetic origin of this excitation (Fig.~\ref{Structure_anisotropy}(f)). 
Fitting the spectra with a Gaussian function (solid line), we found that the elastic peak is resolution-limited with the FWHM around 0.79 meV. However, the FWHM of the inelastic peak around 9.06~meV is about 1.93~meV (the instrumental resolution is 1.2~meV at 9 meV energy transfer), indicating that there might exist a weak spin wave dispersion at the first excited state, CEF-phonon hybridization or disorder. These two scenarios are often considered together with spin-wave dispersion mentioned above, when a CEF peak is not resolution-limited. Also, no other spin excitations were observed at lower energy within the resolution of our setup as expected for an Ising-like spin Hamiltonian~\cite{DSO,TSO}.

The temperature-dependent magnetic susceptibility of NdZnPO is shown in Fig.~\ref{Structure_anisotropy}(g), with the external magnetic field applied in the $ab$-plane ($B\|ab$) and along the $c$-axis ($B\|c$). The Curie-Weiss law does not apply here due to the effect of the CEF excitations at low energy. Instead, significant single-ion anisotropy was observed between the two field directions, which is also evident in the field-dependent magnetization (Fig.~\ref{Structure_anisotropy}(h)). The magnetization measured at 7~T for $B\|c$ is about twice as large as the measured value for $B\|ab$, which is consistent with the calculated single ion ground states with Ising-like moments along the $c$-axis.

\subsection{Zero field antiferromagnetic ground state}
Specific heat measurements down to 0.1 K have been conducted to investigate the low-temperature magnetic properties of NdZnPO. Figure~\ref{CT}(a) displays the zero-field temperature-dependent specific heat $C(T)$ (black circles), while the nonmagnetic isostructural compound YZnPO was employed to estimate the specific heat of lattice contributions (green circles). The magnetic specific heat $C_{\rm M}$ was then obtained by subtracting the lattice contributions (red circles in Fig.~\ref{CT}(b)). A sharp peak was observed at 1.64 K, indicating the establishment of long-range magnetic order. The integrated magnetic entropy exhibits a full entropy of $R\ln2$ around 7 K, confirming a doublet ground state (blue line in Fig.~\ref{CT}(b)). Additionally, the integrated magnetic entropy becomes larger than $R\ln2$ above 7 K. These contributions are likely from the first excited CEF levels. However, we noticed that only about 40\% of the full entropy is released at $T_{\rm N}=1.64$ K. This observation is similar to many other low-dimensional quantum magnets, indicating that the magnetic moments are not completely frozen in the ordered states~\cite{YAO}. In other words, many moments fluctuate even in the magnetically ordered state.

Powder neutron diffraction experiments were performed to determine the magnetic structure of NdZnPO. Figure~\ref{mag_peak}(a,b) presents the powder diffraction patterns measured at 2 K and 0.3 K, respectively. 
The diffraction pattern measured at 2 K can be well refined with the trigonal rhombohedral crystal structure of the space group $R\bar{3}m$ (No.166), consistent with our single-crystal X-ray diffraction results. Only 0.42\% P$_2$O$_5$ has been identified as an additional impurity phase within the powder sample.
However, an additional reflection located at $Q= 0.916~\rm \AA^{-1}$ was observed in the powder neutron diffraction data at 0.3 K, indicating the magnetic origin. 
Using the $\rm K$-search program of the FullProf Suite, we found that this magnetic reflection can be indexed with three equivalent commensurate magnetic propagation vectors: $\mathbf{Q_1}=(1/2,0,1/2)$, $\mathbf{Q_2}=(0,-1/2,1/2)$, and $\mathbf{Q_3}=(-1/2,1/2,1/2)$, as illustrated in Fig.~\ref{mag_peak}(c). 
Figure~\ref{mag_peak}(d) shows powder diffraction patterns taken at several different temperatures between 0.3~K and 2~K. 
Both the magnetic and nuclear peaks are well described by Gaussian functions. The temperature-dependent peak intensity and the FWHM ($\Gamma$) of the peak are presented in Fig.~\ref{mag_peak}(e)-(f). 
The magnetic reflection disappears at the transition temperature around 1.6~K, consistent with our specific heat observations. 
The peak intensity of the (0,0,3) nuclear reflection at $Q= 0.606~\rm \AA^{-1}$ does not change with temperature. 
As the temperature decreases, the $\Gamma$ of the nuclear peak is limited by instrument resolution, maintaining a constant value. In contrast, the $\Gamma$ of the magnetic peak is larger and shows an increase as the transition temperature is approached (Fig.~\ref{mag_peak}(f)).
Note, that at the lowest temperature we measured, $\sim$60~mK, the intensity of the magnetic peak is still not saturated, meaning that the magnetic structure may have some thermal fluctuations or the transition is not finished. 
 

\begin{table}[b]
\renewcommand\arraystretch{1.3}
\begin{ruledtabular}
\caption{Basis vectors of the irreducible representations (IR) for NdZnPO~with magnetic wave-vector $\mathbf{Q_1} =(0.5,0,0.5)$ and Nd site at (0,0,0.38118).}\label{tab_BV}
\centering
\begin{tabular}{cccc}
   IR           & Basis vector   & Nd$_1$(0,0,0.38118)      & Nd$_2$(0,0,0.61882)\\
                &                & $m_{x}$~~$m_y$~~$m_z$ & $m_{x}$~~$m_y$~~$m_z$\\
\hline
     $\Gamma_1$ &  $\psi_1$      &      2~~~~~1~~~~~0    &      -2~~~~~-1~~~~~0\\
                &  $\psi_2$      &      0~~~~~0~~~~~2    &      0~~~~~0~~~~~-2\\
\hline
     $\Gamma_2$ &  $\psi_3$      &      0~~~~~-2~~~~~0   &      0~~~~~-2~~~~~0\\
\hline
     $\Gamma_3$ &  $\psi_4$      &      0~~~~~-2~~~~~0   &      0~~~~~2~~~~~0\\
\hline
     $\Gamma_4$ &  $\psi_5$      &      2~~~~~1~~~~~0    &      2~~~~~1~~~~~0\\
                &  $\psi_6$      &      0~~~~~0~~~~~2    &      0~~~~~0~~~~~2\\
\end{tabular}
\end{ruledtabular}
\end{table}

Representation analysis was applied to explore the potential magnetic structures in NdZnPO. 
Four different irreducible representations (IRs) describe the possible spin configurations for Nd$^{3+}$ ions located at (0,0,0.38118) with the propagation vector $\mathbf{Q_1}=(1/2,0,1/2)$. The basis vectors for IRs $\Gamma_1, \Gamma_2, \Gamma_3$, and $\Gamma_4$ are listed in Table \ref{tab_BV}. Among them, IRs $\Gamma_2$ and $\Gamma_3$ permit the orientation of moments within the $ab$ plane only, contradicting the results of magnetization measurements. IRs $\Gamma_1$ and $\Gamma_4$ allow the magnetic moment to have both in-plane and out-plane components, consistent with the experimental data. Rietveld refinement indicates that IRs $\Gamma_4$ could provide the best fit, yielding R$_{\rm p}$ = 2.15\% and R$_{\rm wp}$ = 3.04\% (Fig.~\ref{mag_peak}(b)). Through the Rietveld refinement, we determined the magnetic structure, which is presented in Fig.~\ref{mag_structure}(a). The ordered magnetic moment is about 1.81(1) $\mu_{\rm B}$/Nd. All the moments are predominantly aligned along the $c$-axis, displaying a small inclination angle of $1.5^\circ$ to the $c$-axis.

Based on the three equivalent magnetic propagation vectors $\mathbf{Q_1}=(1/2,0,1/2)$, $\mathbf{Q_2}=(0,-1/2,1/2)$, and $\mathbf{Q_3}=(-1/2,1/2,1/2)$, there could potentially exist three different magnetic domains for NdZnPO, depicted in Fig.~\ref{mag_structure}(b-d) respectively.
Within these magnetic domains, adjacent Nd$^{3+}$ ions (from two different triangular layers) arrange into zig-zag ferromagnetic (FM) stripes along the [010], [100], and [110] crystallographic directions respectively. These stripes then exhibit AFM alignment along the [100], [010], and [1-10] directions.
Additionally, due to the three-fold rotation symmetry $C_{3}$ for the crystal structure, the magnetic ground state of NdZnPO can be also described by a triple-Q magnetic order which could be understood as a vector sum of the above three magnetic domains, as shown in Fig.~\ref{mag_structure}(e).
A honeycomb ring of six Nd$^{3+}$ ions forms a magnetic unit at the position marked by the black circle.
The total magnetic moment inside the black circle is zero.
However, since both magnetic ground states show the same diffraction patterns, the triple-Q and single-Q (three magnetic domains) magnetic order cannot be distinguished by zero-field neutron diffraction~\cite{Na2Co2TeO6}.
To distinguish these two magnetic structures, future studies, such as single-crystal neutron diffraction experiment with magnetic field, are needed.

\begin{figure*}[hbt!]
\includegraphics[width=1\textwidth]{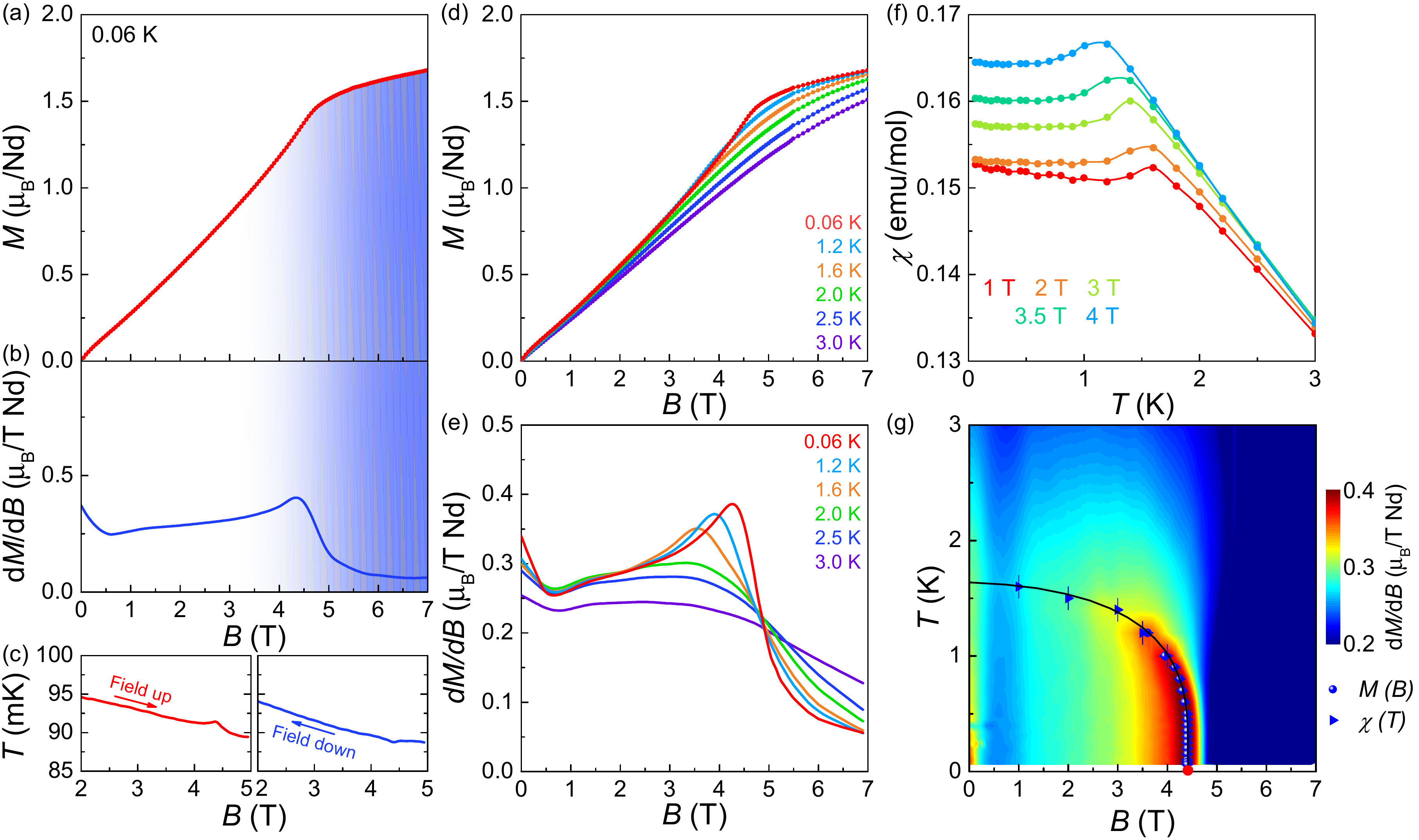}
\caption{(a) Field-dependent magnetization, and (b) magnetic susceptibility d$M/$d$B$, measured at 0.06 K with an applied field along the c axis.
(c) Magnetocaloric effect measured at a constant system temperature of approximately 90 mK by sweeping the magnetic field up (red) and down (blue) with an applied field along the $c$ direction.
(d) Field dependence of magnetization, (e) differential magnetic susceptibility, $dM/dB$, at different temperatures with the field along the $c$-axis.
(f) Temperature dependence of susceptibility $\chi(T)$ at different fields with the field along the $c$-axis.
(g) The field-temperature magnetic phase diagram overlaid on the contour plots of the magnetic susceptibility $dM/dB$ with the field along the $c$ direction. The magnetic phase boundaries were extracted through $M(T)$ and $M(B)$ measurements, with the saturation field $B_{s} = 4.5$ T for $B \parallel c$.}
    \label{magnetization_PD}
\end{figure*}

\begin{figure*}[hbt!]
 \includegraphics[width=1\textwidth]{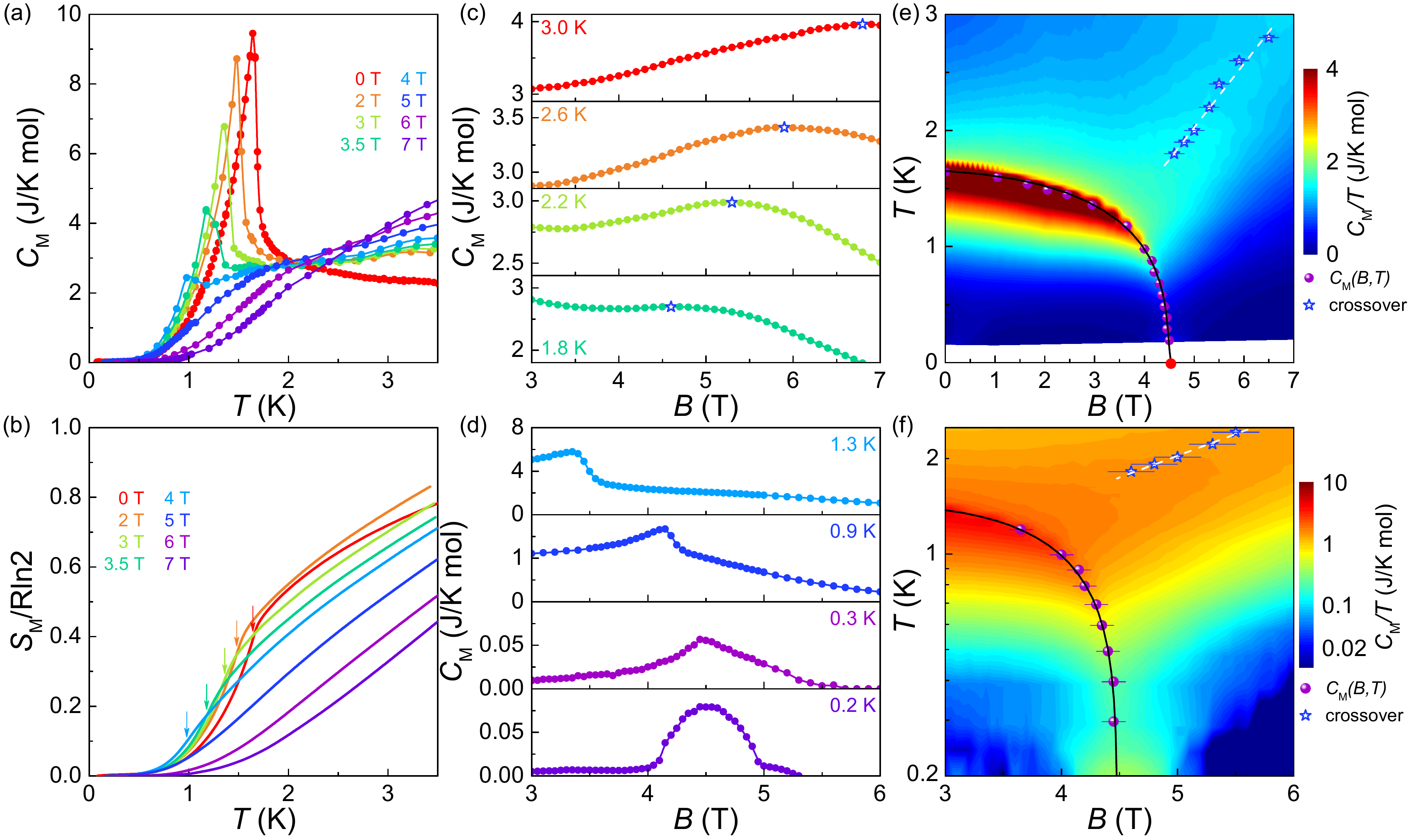}
    \caption{(a) The magnetic specific heat $C_{\rm M}$ and (b) the normalized integrated magnetic entropy $S_{\rm M}$/Rln2 as a function of temperature at different magnetic fields with the field along the c-axis.
(c-d) Field-dependent magnetic specific heat at different temperatures with the field along the c direction.
(e) The field-temperature magnetic phase diagram overlaid on the contour plots of the magnetic heat capacity with the field along the $c$ direction. The magnetic phase boundaries were extracted through $C_{\rm M}(T)$ and $C_{\rm M}(B)$ measurements, with the critical field $B_{\rm c} = 4.5$ T for $B \parallel c$.
(f) A magnified phase diagram around the critical point.}
    \label{HC_PD}
\end{figure*}

\subsection{Field-temperature phase diagram of NdZnPO}

The field-dependent magnetization at various temperatures below $T_{\rm N}$ with $B \parallel c$ was performed as presented in Fig.~\ref{magnetization_PD}. A distinct slope change in the magnetization curve was observed at the base temperature of 60 mK (Fig.~\ref{magnetization_PD}(a)), signifying the transition into a fully polarized state. Accordingly, a peak-like feature was observed at $B_{\rm s}=4.5$ T in the differential susceptibility, $dM/dB$, as indicated in Fig.~\ref{magnetization_PD}(b). To further validate this transition, MCE measurements were conducted. The sample was mounted on a platform made of Sapphire crystal in a high vacuum, and the sample temperature was recorded while sweeping the magnetic field up and down at a rate of 10 Oe/s. The heat was absorbed or released at the phase transition, and kink-like anomalies were observed in the field-temperature curve at the phase boundaries (Fig.~\ref{magnetization_PD}(c)). The critical field determined from the MCE measurements was consistent with the phase boundary established from the differential susceptibility, $dM/dB$. To illustrate the temperature evolution, field-dependent magnetization was systematically measured at different temperatures, as shown in Fig.~\ref{magnetization_PD}(d). The peak-like anomalies in $dM/dB$ observed at the phase boundary gradually became smeared with increasing temperature and finally vanished at temperatures above $T_{\rm N}$ (Fig.~\ref{magnetization_PD}(e)). The temperature dependence of the magnetic susceptibility, $\chi(T)$, is shown in Fig.~\ref{magnetization_PD}(f). Cusp-like anomalies were observed, signaling the establishment of long-range AFM order. The overall phase diagram under the field was summarized in Fig.~\ref{magnetization_PD}(g). The phase boundary extracted from the peak positions in the d$M$/d$B$ and d$\chi$/d$T$ curves is over-plotted on the contour plot of the differential susceptibility. The AFM order at $T_{\rm N}=1.64$~K is gradually suppressed to zero temperature at the critical field $B_{\rm s}=4.5$ T.

Figure~\ref{HC_PD}(a) shows the temperature-dependent specific heat under different magnetic fields. The sharp, peak-like anomalies indicate the establishment of long-range order. These peaks shift towards lower temperatures as the magnetic field increases. Simultaneously, the peak height is suppressed, indicating that less entropy is released at the transition. This suppression is also apparent in the integrated magnetic entropy, as illustrated in Fig.~\ref{HC_PD}(b). The integrated entropy was about half of the full entropy of $R\ln2$ at $T_{\rm N}$ in zero fields. However, as the critical field is approached, the integrated magnetic entropy reduces to only about 10\% of $R\ln2$ at 4.0 T.
This hints that the residual entropy might be released at temperatures beyond the transition, possibly implying the persistence of strong short-range correlations near the critical field. However, confirmation of magnetic diffuse scattering at temperatures above the phase transition would require future single-crystal neutron scattering experiments.

Field-dependent specific heat was also measured at various temperatures, as shown in Fig.~\ref{HC_PD}(c) and (d). A broad Schottky-like anomaly was observed at temperatures above $T_{\rm N}$, with the peak positions extending linearly to lower temperatures as the field decreases (blue stars in Fig.~\ref{HC_PD}(c) and (e)). 
The Schottky anomaly in the specific heat is a general feature of two-level systems. In the case of NdZnPO, the doublet ground state splits in fields due to the Zeeman effect. Typically, for a system with two discrete flat levels, the peak value in the specific heat of the Schottky anomaly is fixed to be close to 3.6~J/mol K. 
However, we noticed that the measured peak values in specific heat show a significant deviation as the temperature and field decrease. This indicates that weak dispersion persists in the system, and these discrete levels are not completely flat. This weak dispersion becomes increasingly significant as the Zeeman gap closes when $B\rightarrow B_{\rm s}$. As the measuring temperature was lower than $T_{\rm N}$, the broad Schottky-like anomaly disappears, and instead, a sharp peak-like feature appears at the critical fields (Fig.~\ref{HC_PD}(d)). 
The transition gradually approaches the saturation field $B_{\rm s}=4.5$ T as the temperature further decreases. The contour plot of the magnetic phase diagram established by the specific heat measurements is presented in Fig.~\ref{HC_PD}(e). 
The phase boundaries extracted from the specific heat and the magnetization are consistent with each other.

Additionally, a `peculiar' feature appears below about 0.3 K. We noticed that the sharp peak-like feature gradually evolves into a broad maximum near the critical field at the base temperature (the bottommost curve in Fig.~\ref{HC_PD}(d)). This broad maximum-like feature extends over a wide field range from about 4 T to 5 T. This is also evident in the contour plot of the temperature-divided specific heat $C_{\rm M}/T$, as shown in Fig.~\ref{HC_PD}(f). At first glance, it seems like a new `dome'-shaped phase was induced at the critical field.
However, our observations hint that this additional contribution might be attributed to the nuclear Schottky effect. 
Upon thorough examination, we noticed that this specific field range exactly corresponds to the maximum of the differential susceptibility $dM/dB$~\cite{ADR}, as depicted in Fig.~\ref{magnetization_PD}(e) and (g). 
As a result, the internal field, which correlates with the differential susceptibility, is significantly enhanced within this field range.
Consequently, it appears that the nuclear moments are polarized by the internal field rather than the external field. 
This explains why this contribution decreases again when the field exceeds 5~T.
Similar behavior has been observed in other rare-earth-based compounds near the saturation field, as documented in previous studies~\cite{TSO}. Furthermore, this effect holds potential for the use of nuclear adiabatic demagnetization cooling techniques under moderate external fields in future applications~\cite{ADR}.

\section{Conclusion and discussion}
To conclude, we successfully synthesized high-quality single crystals of NdZnPO, which adopt well-separated bilayer triangular lattices composed of magnetic Nd$^{3+}$ ions. Thermal property measurements, including magnetization and specific heat, have been performed. The experimental results, together with the CEF analysis, revealed a doublet ground state with Ising-like moments aligned along the $c$-axis. In zero magnetic field, a long-range AFM order was established at $T_{\rm N}=1.64$ K. The magnetic structure has been refined by the powder neutron diffraction measurements. These Ising moments form into FM stripes in the $ab$-plane, with three equivalent magnetic propagation vectors ($\mathbf{Q_1}=(1/2,0,1/2)$, $\mathbf{Q_2}=(0,-1/2,1/2)$, and $\mathbf{Q_3}=(-1/2,1/2,1/2)$). With the external field applied along the easy $c$-axis, this long-range magnetic order was gradually suppressed with a saturation field $B_{\rm s}=4.5$~T. Additionally, we also noticed that the nuclear Schottky contribution was greatly enhanced by the internal field arising from the 4$f$-electrons, resulting in a dome-shaped anomaly in specific heat centered around $B_{\rm s}$.

Although the magnetic structure and the magnetic phase diagram of NdZnPO have been established with detailed thermodynamic characterizations, several key issues still need to be clarified in the future. As we mentioned in Section~\ref{structure}, NdZnPO hosts well-separated 2D bilayer triangular lattices with the nearest neighbor distances between these bilayers larger than 7~\AA. In this case, even the long-range dipolar interactions between these Ising-like Nd$^{3+}$ spins in different bilayers are negligibly small (around 10~mK). 
Moreover, considering the considerable separation between localized 4$f$-electrons, it becomes challenging to envision the exchange interaction playing a significant role. Besides, the Ruderman-Kittel-Kasuya-Yosida interactions are absent since NdZnPO is a transparent insulator. According to the Mermin-Wanger theorem, it is hard to imagine that the long-range magnetic order can build up at temperature $T_{\rm N}=1.64$~K for such well separated 2D bilayer trianglular lattices? An likely senario is that, the strong Ising-like anisotropy has helped to stabilize the magnetic ground state. In addition, the Ising-like single ion anisotropy results in a large saturation moment, and thus the long-range dipolar intra-plane interactions are also enhanced. This long-range interaction also makes the Mermin-Wanger theorem fail in the case of NdZnPO. We also noticed that a long-range magnetic order was also established at 1.3~K in the compound YbOCl with similar bilayer triangular lattices~\cite{YOC1,YOC2}. These resembling phenomena in these two systems might indicate that the magnetic order may be stabilized by similar mechanisms.

Additionally, the magnetic dipolar interaction calculation between near neighbors indicates that these moments are AFM coupled ($J^\prime=0.196~K$) in the single triangular layer, while they are FM coupled ($J=-0.102~K$) between these bilayers. Alternatively, this is equivalent to a honeycomb lattice with FM nearest-neighbor interactions ($J$) and AFM next-nearest-neighbor interactions ($J^\prime$). For Ising moments in this honeycomb lattice with next-nearest-neighbor interactions, strong quantum fluctuations are expected due to geometric spin frustration. Assuming the static magnetic structure could be stabilized by a delicate balance of these nearest-neighbor and next-nearest-neighbor interactions, it is still likely that the ordered moment would be greatly reduced due to the strong quantum fluctuations.
However, in contrast to our expectation, the ordered moments $1.81~\mu_{\rm B}$/Nd refined from the powder neutron diffraction experiments are very close to the saturation moment $1.68~\mu_{\rm B}$/Nd at 7~T observed in the magnetization measurements. This suggests that most of the spin frustration has been released in the magnetic ordering state. Besides, fractional states such as 1/3 plateau phases are usually induced under fields for triangular lattices with Ising-like moments. However, we did not find any field-induced plateau-like phases down to the base temperature in NdZnPO. To explain these observations and to check if Kitaev-like physics could emerge in this system, further investigations such as single-crystal neutron scattering experiments are still needed.

\begin{acknowledgments}
We would like to thank Shu Guo for the help with the single crystal growth, Nikita Andriushin for assisting in the neutron scattering experiment, and Zhentao Wang for useful discussions. 

The research was supported by the National Key Research and Development Program of China (Grant No.~2021YFA1400400), the National Natural Science Foundation of China (Grants No.~12134020, No.~12374146, No.~12104255, No.~12005243 and No.~11974157), the Open Fund of the China Spallation Neutron Source Songshan Lake Science City (Grant No.~KFKT2023A06), and the Guangdong Basic and Applied Basic Research Foundation (Grant No.~2021B1515120015 and No.~ 2022B1515120014).
Part of this work is based on experiments performed at the Swiss spallation neutron source SINQ, Paul Scherrer Institute, Villigen, Switzerland.


\end{acknowledgments}

\end{document}